\documentclass[pre,twocolumn,,floatfix]{revtex4}
\usepackage{graphicx,citesort,psfrag,amsmath,amssymb}
\usepackage[dvips]{color}

\newcommand{\avg}[1]{\left\langle #1\right\rangle}
\newcommand{\abs}[1]{\left\lvert #1\right\rvert}

\newcommand{\gex}{P}
\newcommand{\hex}{H}
\newcommand{\gst}{P_{\text{st}}}

\newcommand{\cst}{c_{\text{st}}}

\newcommand{\cex}{n}
\newcommand{\nef}{N_{\text{e}}}

\newcommand{\pushed}{\textit{pushed}}
\newcommand{\pulled}{\textit{pulled}}

\begin{document}
\bibliographystyle{apsrev}

\title{Gene surfing in expanding populations}

\author{Oskar Hallatschek\footnote{To whom correspondence should be addressed.
    E-mail: ohallats@physics.harvard.edu} } \author{David R.~Nelson}
\affiliation{Lyman Laboratory of Physics, Harvard University, Cambridge,
  Massachusetts 02138, USA }

\date{\today}

\begin{abstract}
  Spatially resolved genetic data is increasingly used to reconstruct the migrational
  history of species.  To assist such inference, we study, by means of simulations
  and analytical methods, the dynamics of neutral gene frequencies in a population
  undergoing a continual range expansion in one dimension.  During such a
  colonization period, lineages can fix at the wave front by means of a ``surfing''
  mechanism [Edmonds C.A., Lillie A.S. \& Cavalli-Sforza L.L.  (2004) Proc Natl Acad
  Sci USA 101: 975-979].  We quantify this phenomenon in terms of (i) the spatial
  distribution of lineages that reach fixation and, closely related, (ii) the
  continual loss of genetic diversity (heterozygosity) at the wave front,
  characterizing the approach to fixation. Our simulations show that an
  \textit{effective population size} can be assigned to the wave that controls the
  (observable) gradient in heterozygosity left behind the colonization process. This
  effective population size is markedly higher in {\pushed} waves than in {\pulled}
  waves, and increases only sub-linearly with deme size.  To explain these and other
  findings, we develop a versatile analytical approach, based on the physics of
  reaction-diffusion systems, that yields simple predictions for any deterministic
  population dynamics.
\end{abstract}

\maketitle

Population expansions in space are common events in the evolutionary
history of many
species~\cite{templeton2002oaa,RosenbergPWCKZF03,RamachandranDRRFC05,cavallisforza93,phillips2006iae,Hewitt00,currat06}
and have a profound effect on their genealogy.  It is widely appreciated that any
range expansion leads to a reduction of genetic diversity (``Founder Effect'')
because the gene pool for the new habitat is provided only by a small number of
individuals, which happen to arrive in the unexplored territory first. In many
species, the genetic footprints of these \textit{pioneers} are still recognizable
today and provide information about the migrational history of the species. For
instance, a frequently observed south-north gradient in genetic diversity (``southern
richness to northern purity''~\cite{RefWorks:32}) on the northern hemisphere is
thought to reflect the range expansions induced by the glacial cycles. In the case of
humans, the genetic diversity decreases essentially linearly with increasing
geographic distance from Africa~\cite{RosenbergPWCKZF03,RamachandranDRRFC05}, which
is indicative of the human migration out of Africa. It is hoped~\cite{RefWorks:19},
that the observed patterns of neutral genetic diversity can be used to infer details
of the corresponding colonization pathways.

Such an inference requires an understanding of how a colonization process generates a
gradient in genetic diversity, and which parameters chiefly control the magnitude of
this gradient. Traditional models of population genetics~\cite{RefWorks:33}, which
mainly focus on populations of constant size and distribution, apply to periods
before and after a range expansion has occurred, when the population is at
demographic equilibrium.  However, the spatio-temporal dynamics in the transition
period, on which we focus in this article, is less amenable to the standard
analytical tools of population genetics, and has been so far studied mostly by means
of simulations~\cite{nichols94,AusterlitzG03,EdmondsLC04,KlopfsteinCE06,LiuPMB06}.
An analytical understanding is available only for a linear stepping stone model in
which demes (lattice sites) are colonized one after the other, following
deterministic logistic growth~\cite{AusterlitzJGG97} or
instantaneously~\cite{LeCorreK98}, in terms of recurrence relations.

\begin{figure*}
   \centering
  \centerline{\includegraphics{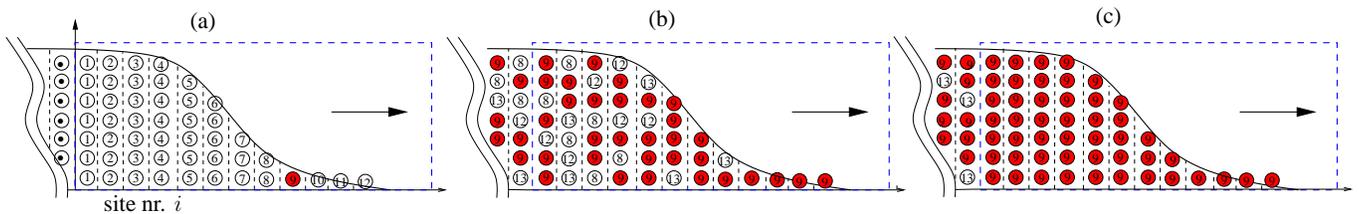}}
  \caption{ Illustration of gene surfing by means of three consecutive
    snapshots of the genetic composition at the edge of an expanding population (in
    one spatial dimension). (a) A neutral {\it red} mutant arises at the wave front.
    (b) After some time, the genetic make-up at the wave front is drastically changed
    due to random number fluctuations and it is apparent that descendents of {\it
      red} will take over the wave front. (c) Fixation in the co-moving frame of
    descendents of {\it red}.  Numbers in these sketches represent ``inheritable''
    labels that are used in our simulations to trace back the spatial origin of
    individuals in the wave front.  In this example, descendents of {\it red} are
    associated with position ``9'' in the co-moving frame.  The dashed blue frame
    indicates the co-moving simulation box.}
  \label{fig:labeled-wave}
\end{figure*}

Recent computer studies suggest that the neutral genetic patterns created by a
propagating population wave might be understood in terms of the mechanism of ``gene
surfing''~\cite{EdmondsLC04,KlopfsteinCE06}: As compared to individuals in the wake,
the pioneers at the colonization front are much more successful in passing their
genes on to future generations, not only because their reproduction is unhampered by
limited resources but also because their progeny start out from a good position to
keep up with the wave front (by means of mere diffusion).  The offspring of pioneers
thus have a tendency to become pioneers of the next generation, such that they, too,
enjoy abundant resources, just like their ancestors. Therefore, pioneer genes
have a good chance to be carried along with the wave front and attain high
frequencies, as if they ``surf'' on the wave.  Thus, the descendents of an
individual sampled from the tip of the wave have a finite probability to take over
the wave front. In this case of ``successful surfing'', further colonization will
produce only descendents of the relevant pioneer because the wave front has been
``fixed''. The process of fixation at the front of a one-dimensional population wave
is illustrated in Fig~\ref{fig:labeled-wave}.

The present study hinges on the question as to \textit{where} lineages that reach
fixation originate within the wave front. Clearly, the probability of successful
surfing must increase with the proximity to the edge of the
wave~\cite{KlopfsteinCE06}. On the other hand, more surfing \textit{attempts}
originate from the bulk of the wave where the population density is larger. We show
that, due to this tradeoff, the origins of successful lineages have a bell-like
distribution inside the wave front. Furthermore, this ancestral probability
distribution, together with the population-density profile of the population wave
itself, is found to control the observable gradients in genetic diversity. The
genetic pattern directly behind the moving colonization front turns out to mimic that
of a small well-mixed (\emph{panmictic}) population.  The {\it effective} size $N_e$
of this population ``bottleneck'' is shown to be {\it smaller} than the typical
number of individuals in the colonization front and very sensitive to the growth
conditions in the very tip of the front. Colonization fronts in which individuals
need to be accompanied by others in order to grow ({\it Allee} effect~\cite{Allee31})
have a much larger effective population size than those in which individuals grow
even if they are isolated from the rest.

The outline of the paper is as follows. We first introduce a stochastic computer
model that we use to generate both {\pulled} and {\pushed} one-dimensional
colonization waves.  Tracer experiments within this model are then used to reveal the
probability distribution of successful surfers and the decrease of genetic
differentiation at the colonization front. Our succeeding theoretical treatment
reveals to what extent both measures are related, and how they can be predicted for
continuous models with quasi-stationary demography.  After a comparison between
theory and simulations, we discuss the significance of our results in the light of
inferring past range expansions from spatially resolved genetic data.

\section{Simulations}
\label{sec:simulations}
Most models for range expansions can be classified as describing {\pulled} or
{\pushed} population fronts~\cite{van03,panja04}. The distinction between the two
cases corresponds to a difference in behavior.  Suppose individuals need to be in
proximity to other individuals in order to grow in number (Allee
effect~\cite{Allee31}). The presence of conspecifics can be beneficial due to
numerous factors, such as predator dilution, antipredator vigilance, reduction of
inbreeding and many others~\cite{courchamp99}. Then, the individuals in the very tip
of the front do not count so much, because the rate of reproduction decreases when
the number density becomes too small.  Consequently, the front is {\pushed} in the
sense, that its time--evolution is determined by the behavior of an ensemble of
individuals in the boundary region.  On the other hand, a population in which an
individual reproduces, even if it is completely isolated from the rest, will be
``spearheaded'' by these front individuals.  These {\pulled} fronts are responsive to
small changes in the frontier and, therefore, are prone to large
fluctuations~\cite{panja04}.

One might suspect that the genetic pattern left behind a population wave should
reflect whether the colonization process is controlled by a small or large number of
individuals. Hence, we have set up a computer model that allows us to investigate the
surfing dynamics for both classes of waves.

\subsection{Population dynamics} The population is distributed on a one dimensional
lattice, whose sites (demes) can carry at most $N$ individuals. The algorithm
effectively treats individuals ($\bullet$) and vacancies ($\circ$) as two types of
particles, whose numbers must sum to $N$ at each site. A computational time step
consists of two parts: (i) a migration event, in which a randomly chosen particle
exchanges place with a particle from a neighboring site. This step is independent of
the involved particle types.  (ii) A duplication {\it attempt}: Two particles are
randomly chosen (with replacement) from the same lattice site. A duplicate of the
first one replaces the second one ($1\textsuperscript{st}\to 2\textsuperscript{nd}$)
with probabilities based on their identities: proposed replacements $\circ\rightarrow
\circ$, $\bullet\rightarrow\bullet$ and $\bullet\rightarrow\circ$ (growth) are
realized with probability $1$, whereas $\circ\rightarrow\bullet$ (death) is carried
out only with probability $1-s$, depending on a growth parameter $0<s<1$.  This
asymmetry controls the effective local growth advantage of $\bullet$ over $\circ$.

In terms of individuals and vacancies instead of particles, we see that our model
describes migration and local logistic growth of a population distributed over demes
with carrying capacity $N$. Starting with a step-function initial condition, the
simulation generates an expanding {\pulled} population wave. The above algorithm
represents a discretized version~\cite{RefWorks:30} of the stochastic
Fisher-Kolmogorov equation~\cite{fisher37} with a Moran-type of breeding
scheme~\cite{RefWorks:33}.  To generate {\pushed} waves as well, we extend our model
by the following rule: In demes in which the number of individuals falls to $N_c$ or
below, we set their effective linear growth rate $s$ to zero.  This represents, for
$N_c>0$, a simple version of the above mentioned Allee effect of a reduced growth
rate when the population density is too small.

\subsection{Tracer dynamics} Tracer experiments within this computer model allow us
to extract the genealogies of front individuals. After the population had enough time
to relax into its propagating equilibrium state, all individuals are labeled
according to their current position $i\in\{1\dots n\}$ within the simulation box of
length $n$, see Fig.~\ref{fig:labeled-wave}a. These labels are henceforth inherited
by the descendents, which thereby carry information about the spatial position of
their ancestors. The randomness in the reproduction and migration processes (genetic
drift) during the succeeding dynamics inevitably leads to a reduction in the
diversity of labels present in the simulation box, see Fig.~\ref{fig:labeled-wave}b.
Labels are lost due to either extinction or because they cannot keep up with the
simulation box, which {\it follows} the propagating wave front\footnote{ The
  simulation box is forced to move with the wave front such that it always contains
  less than a given large number $M$ of individuals. We usually set $M=45\times N$
  individuals, where $N$ is the deme size.  The size $n$ of the simulation box had to
  be chosen so that it contained the entire front up to the foremost individual. One
  hundred lattice sites usually were sufficient.}.

In our simulation, the gradual loss of diversity of labels at the wave front is
measured by the quantity
\begin{equation}
  \label{eq:Phi}
  \hex(t)=\sum_{i=1}^{n}p_i(t)\left[1-p_i(t)\right] \;,
\end{equation}
which depends on the frequency $p_i(t)$ of label $i$ at time $t$ after the wave has
been labeled.  $\hex(t)$ represents the time-dependent probability that two
individuals, randomly chosen from the bounded simulation box, carry different labels.
Provided that mutations are negligible on the time-scale of the range expansion, we
may think of our inheritable labels as being neutral genes at one particular locus
(alleles).  We may thus identify $\hex(t)$ with the probability that two alleles
randomly chosen from the front region are different conditional on the well-mixed
labeling state at $t=0$ imposed by our simulation.  Hence, we refer to $\hex(t)$ as
the time-dependent \textit{expected heterozygosity}~\cite{RefWorks:33} at the wave
front~\footnote{If the average wave velocity is $v$, then the heterozygosity at
  position $x$ (in the non-moving frame) as the wave front passes through is given by
  $H(x/v)$.  }.

The perpetual loss of labels  in our model without mutations eventually leads to the
fixation of one label in the simulation box, see Fig.~\ref{fig:labeled-wave}c. The
value of this label indicates the origin \emph{within the co-moving frame} of this
successful ``surfer''.  It contributes one data point to the spatial distribution
$\gex_i$ of individuals whose descendents came to fixation.  After fixation, the
algorithm proceeds with the next labeling event.

\subsection{Results} The parameters of our computer models are the deme size $N$,
i.e.  the maximal number of individuals per lattice site, the linear growth rate $s$
per generation, and the critical occupation number $N_c$, below which the growth rate
drops to zero (Allee effect). In our simulations, we set $s=0.1$ throughout, and
determine, for varying $N$ and $N_c$, the averages of the ancestral distribution
function $\avg{\gex_i}$, the scaled occupation number $\avg{\cex_i}/N$, both being
functions of the lattice site $i$ in the co-moving frame, and the time-dependent
probability of non-identity, $\avg{\hex(t)}$. Here, angle brackets indicate that the
enclosed quantities have been averaged in time, i.e. over many fixation events, {\it
  and} over multiple realizations of the same computer experiment\footnote{For the
  smallest deme sizes $N=30$, we carried out $10$ realizations each measuring $10^5$
  fixation processes. We have less statistics for larger deme sizes because of the
  larger number of degrees of freedom and fixation times. For $N=36100$, we ran $30$
  simulations and measured $500$ fixation processes.}.

\begin{figure*}
  \includegraphics{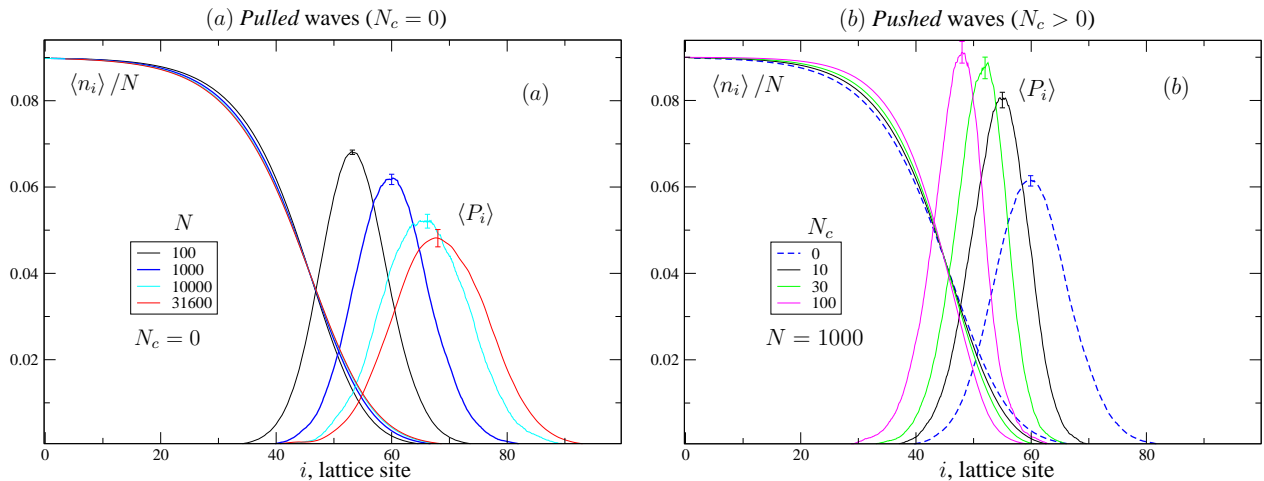}
    \caption{ Measured distributions $\avg{\gex_i}$ (bell-curves) of
      ``successful surfers'' together with the normalized occupation numbers
      $\avg{\cex_i}/N$ (sigmoidal curves; scaled along the vertical axis to fit the
      figure) as a function of the site number $i$ in the co-moving frame; $(a)$ for
      {\pulled} waves ($N_c=0$) with varying deme sizes $N$; $(b)$ for various
      {\pushed} waves ($N_c> 0$) with deme size $N=1000$ compared to the
      corresponding {\pulled} wave (dashed blue lines), which is also present in
      (a).}
  \label{fig:exple}
\end{figure*}

Figure \ref{fig:exple} illustrates the relation between the front profiles
$\avg{\cex_i}/N$ and the ancestral distribution $\avg{\gex_i}$ in the co-moving
frame. Whereas the wave profiles have the familiar sigmoidal shapes of
reaction-diffusion waves~\cite{van03,panja04}, the ancestral distribution functions
are bell-curves with most of its support beyond the inflection point of the wave
front.  The fact that $\avg{\gex_i}$ has a maximum inside the wave front reflects a
tradeoff, mentioned earlier, between a larger fixation probability in the tip of the
wave versus a larger number of surfing {\it attempts} originating from the bulk. 
Notice from Fig.~\ref{fig:exple}a that, for increasing deme size, the distribution
becomes wider and shifts further into the tip of the wave, which is in contrast to
the almost $N-$independent scaled wave profiles.  Fig.~\ref{fig:exple}b shows that
the opposite effect is caused by increasing the cutoff value $N_c$, which changes the
type of the wave from {\pulled} to {\pushed}.

\begin{figure*}
  \includegraphics{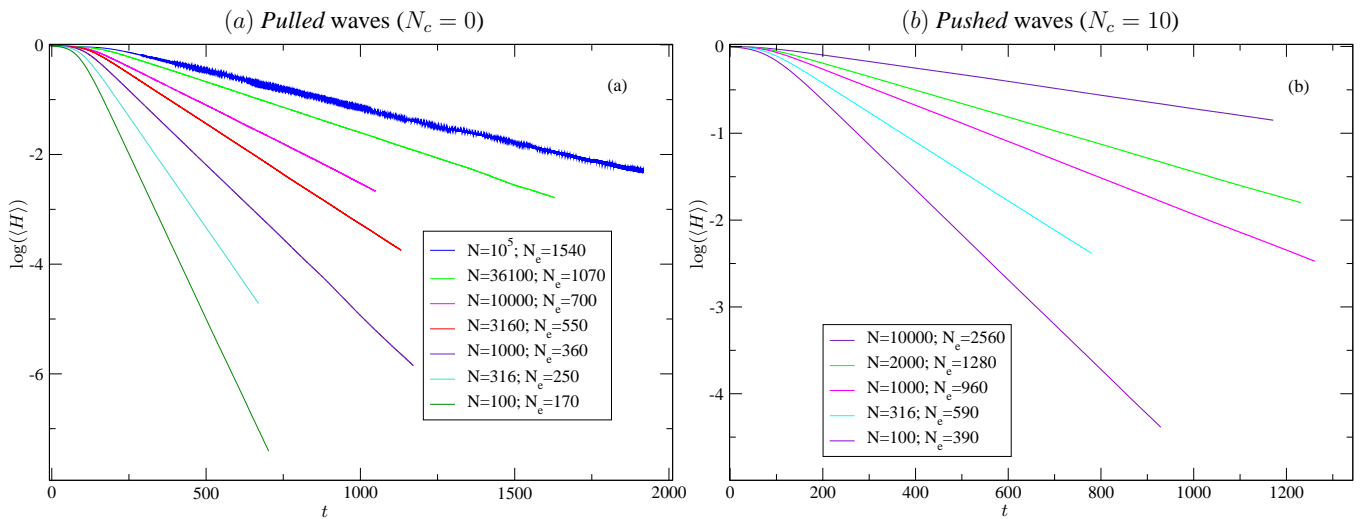}
  \caption{The decay of genetic diversity, $\avg{\hex(t)}$, with time
    (in units of generations) on a log-linear scale for varying deme sizes; (a) for
    {\pulled} waves ($N_c=0$); (b) for {\pushed} waves with $N_c=10$. Both cases
    show, that an asymptotic exponential decay of $\avg{\hex(t)}$ is reached after an
    initial transient where the decay is weak.  The duration of this transient is
    dependent on the size of the simulation box: The larger the simulation box, the
    larger the time until the exponential decay is approached. The asymptotic
    exponential decay rate, however, has been checked to approach a constant for a
    sufficiently large box size. This exponential decay rate is therefore
    well-defined and can be used to characterize the decrease of genetic diversity at
    the wave front.  By analogy with panmictic populations, in which the
    heterozygosity decays exponentially with rate $2/N$ (Moran model), it is
    convenient to express the decay rate as $2/N_e$, i.e., in terms of an {\it
      effective population size} $N_e$, which is noted in the legends, and plotted in
    Fig.~\ref{fig:log-log-effpopsizes}.}
  \label{fig:log-lin-heterozygosity}
\end{figure*}

Next, we measured the temporal decay of the heterozygosity $H(t)$, defined in
Eq.~(\ref{eq:Phi}).  In Fig.~\ref{fig:log-lin-heterozygosity}, time-traces of $\hex(t)$ are
depicted for various parameters and show an exponential decay after an initial
transient
.  This allows us to characterize the strength of genetic drift at the wave front by
a single number, the (asymptotic) exponential decay rate,
$-\partial_t\log\avg{\hex(t)}$, which can be extracted from logarithmic plots of
$\avg{H(t)}$. By analogy with well-mixed (panmictic) populations, in which the
heterozygosity decays exponentially with rate $2/N$ (Moran model\cite{RefWorks:33}),
it is convenient to express the decay rate by $2/N_e$, in terms of an {\it effective
  population size} $N_e$. The theoretical part below will further clarify to what
extent the genetic diversity at the wave front mimics that of a population
``bottleneck'' of constant size $N_e$.

\begin{figure}
 \includegraphics{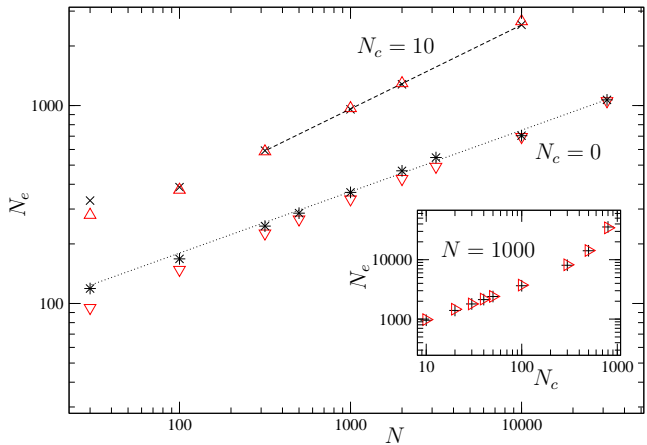}
 \caption{ The measured effective population size $N_e$ as a function of deme size
   $N$ on a log-log scale for {\pulled} waves ($N_c=0$, asterisks) and {\pushed}
   waves ($N_c=10$, crosses).  The dashed and dotted lines have slope $.30$ and
   $.42$, respectively, i.e., significantly smaller than $1$.  Triangles represent the
   effective population sizes as inferred from the strong-migration approximation,
   Eq.~(\ref{eq:effective-popsize}), using the \textit{measured}
   $\avg{\gex}$--distribution and population profiles.  The inset shows the behavior
   of $N_e$ for varying cutoff-value $N_c$ and fixed deme size $N=1000$, again, on a
   log-log scale.}
  \label{fig:log-log-effpopsizes}
\end{figure}

Figure~\ref{fig:log-log-effpopsizes} depicts $N_e$ as a function of the deme size $N$
on a double logarithmic scale for $N_c=0$ and $N_c=10$. Naively, one might expect
$N_e$ to be, roughly, the characteristic number of individuals in the width of the
wave front, since these individuals contribute (by growing) to the advance of the
wave. Thus, a linear relationship between deme and effective population size would
not be surprising. In contrast, we find that $N_e$ increases much slower than
linearly with increasing deme size.  Furthermore, the effective population size turns
out to be very sensitive to the presence of an Allee effect ($N_c>0$), which has the
ability strongly increase the effective population size.  This point is illustrated,
in particular, by the inset of Fig.~\ref{fig:log-log-effpopsizes} which depicts the
effective population size $N_e$ in a simulation of fixed deme size ($N=1000$) and
varying strength of the Allee effect ($10<N_c<500$).  Qualitatively this phenomenon
may be explained with the {\pushed} nature of these waves.  An Allee effect shifts
the distribution $P_i$ of successful surfers away from the tip towards the wake of
the wave (see Fig.~\ref{fig:exple}b) and hence increases the gene pool from which the
next generation of pioneers is sampled. This argument indicates a close relation
between the $N_e$ and $P_i$, which also emerges explicitely in the theoretical
analysis below.

\section{Theory}
\label{sec:formalism}
The following employs a continuous reaction diffusion approach to establish a
theoretical basis for the relation between the neutral genetic diversity and the
population dynamics in non-equilibrium situations like range expansions.  It will
help us to reconcile the somewhat surprising response of our simulations to parameter
changes (deme size and Allee effect).  Note from Fig.~\ref{fig:exple} that the
changes in the ancestral distribution are dramatic, while the changes in the
population profile itself are quite modest.  Results obtained from our approximation
scheme are tested by direct comparison of simulations and theory.

\subsection{Gene surfing} In our simulations, as well as in many other
models of range expansions, a propagating population wave results from the
combination of random short-range migration and logistic local growth. In the
continuum limit, a general coarse-grained continuum description of such a
reaction-diffusion system of a single species is given by
\begin{equation}
  \label{eq:comoving-Rdiff}
  \partial_t c(x,t)=D\partial_x^2 c(x,t)+v\partial_x c(x,t)+K(x,t)
\end{equation}
formulated in the frame co-moving with velocity $v$, where $c(x,t)$ represents the
density of individuals at location $x$ at time $t$ and $D$ is a diffusivity. The
first two terms on the right hand side represent the \emph{conservative} part of the
population dynamics, for which we make the usual diffusion
assumptions~\cite{vankampen01}. The reaction term $K(x,t)$ accounts for both
deterministic and stochastic fluctuations in the number of individuals due to birth
and death processes, and typically involves non--linearities such as a logistic
interaction between individuals as well as noise caused by number fluctuations. For
instance, our computer model with $N_c=0$ maps, in the continuum
limit~\cite{RefWorks:30}, to the stochastic Fisher equation, for which $K(x,t)=s c
(c_\infty-c)+\epsilon \sqrt{c (c_\infty-c)}\eta$, where $\eta(x,t)$ is a Gaussian
white noise process in space and time, $c_\infty\propto N$ is the carrying capacity
and $\epsilon\propto\sqrt{N}$ sets the strength of the noise. We would like to
stress, however, that the following analysis does not rely on a particular form of
$K$. Therefore, we leave the reaction term unspecified.


As in our tracer experiments, let us assume that inheritable labels, representative
of neutral genes, are attached to individuals within the population and ask: Given
Eq.~(\ref{eq:comoving-Rdiff}) is a proper description of the population dynamics, to
what extent is the dynamics of these labels determined? To answer this question, it
is convenient to adopt a retrospective view on the tracer dynamics.  Imagine
following the ancestral line of a single label located at $x$ \emph{backwards} in
time to explore which spatial route its ancestors took.  This backward--dynamics of a
single line of descent will show drift and diffusion \emph{only}; any reaction is
absent because among all the individuals living at some earlier time there must be
exactly one ancestor from which the chosen label has descended from.  We may thus
describe the ancestral process of a single lineage by the probability density
$G(\xi,\tau|x,t)$ that a label presently, at time $t$ and located at $x$, has
descended from an ancestor that lived at $\xi$ at the earlier time $\tau$. In this
context, it is natural to choose the time as increasing towards the past, $\tau>t$,
and to consider $(\xi,\tau)$ and $(x,t)$ as \emph{final} and \emph{initial} state of
the ancestral trajectory, respectively. With this convention, the distribution $G$
satisfies the initial condition $G(\xi,t|x,t)=\delta(x-\xi)$, where $\delta(x)$ is
the Dirac delta function, and is normalized with respect to $\xi$, $\int
G(\xi,\tau|x,t) \,d\xi=1$.

Since $G(\xi,\tau|x,t)$ as function of $\xi$ and $\tau$ is a probability distribution
function generated by a diffusion process that is continuous in space and time, we
expect its dynamics to be described by a generalized diffusion equation
(Fokker--Planck equation~\cite{vankampen01}). Indeed, in the Appendix \ref{app:a} we show that,
$G(x,t|\xi,\tau)$ obeys
\begin{eqnarray}
  \label{eq:FPE}
  \partial_\tau G( \xi,\tau|x,t)&=&-\partial_\xi
  J(\xi,\tau|x,t) \\
  J(\xi,\tau|x,t)&\equiv &-D\partial_\xi G+\left\{v+2 D\partial_\xi \ln[c(\xi,\tau)] \right\} G \nonumber  \;,
\end{eqnarray}
where all derivatives are taken with respect to the ancestral coordinates
$(\xi,\tau)$.  The drift term in Eq.~(\ref{eq:FPE}) has two antagonistic parts. The
first term, $v$, tends to push the lineage into the tip of the wave, and is simply a
consequence of the moving frame of reference.  The second term proportional to
\emph{twice} the gradient of the logarithm of the density is somewhat unusual. It
accounts for the purely ``entropical'' fact that, since there is a forward--time flux
of individuals diffusing from regions of high density to regions of low density, an
ancestral line tends to drift \emph{into} the wake of the wave where the density is
higher.\footnote{ In a deterministic analysis of the simulations in
  Ref.~\cite{EdmondsLC04}, the Kolmogorov-backward equation~\cite{vankampen01}
  associated with Eq.~(\ref{eq:FPE}) has recently been obtained by Vlad et
  al.~\cite{VladCR04}.}


Our computer experiments measure the spatial distribution $\gex$ of the individuals
whose descendents came to fixation. This information is encoded in the long-time
behavior of $G$
\begin{equation}
  \label{eq:surfing-theory}
  \gex(\xi,\tau)=\lim_{t\to-\infty}G(\xi,\tau|x,t)\;,
\end{equation}
because it represents the probability that, in the far future ($t\to-\infty$, in our
notation) when the population is fixed, an individual of the extant lineage has
descended from an individual who lived at location $\xi$ of the co-moving frame at
time $\tau$.  Equation (\ref{eq:surfing-theory}) has to be independent of $x$ if
fixation occurs: For $t\to-\infty$, all individuals irrespective of their position
$x$ must have descended from the same ancestor and, thus, from the same location at
the earlier time $\tau$.

In principle, it is thus possible to relate the ancestral distribution to the
population dynamics by solving Eq.~(\ref{eq:FPE}) in the long-time
limit. 
Unfortunately, this task is usually difficult to achieve analytically because the
number fluctuations in the density $c(\xi,\tau)$ of the total population add noise to
the drift term in Eq.~(\ref{eq:FPE}). As is customary in many spatially explicit
models of population genetics, let us suppose, however, that rules of ``strict
density regulation''~\cite{BARTON:W::349:p49-59:1995}
are imposed in order to guarantee a stationary demography, so that in the co-moving
frame,
\begin{equation}
  \label{eq:stationarity}
  c(x,t)\approx\cst(x) \;.
\end{equation} 
Even though real systems and our discrete particle simulations exhibit density
fluctuations even in equilibrium, we take Eq.~(\ref{eq:stationarity}) as a first
approximation in cases where the total number of particles is large enough, such that
the relative magnitude of the density fluctuations is small (law of large numbers).
We will call assumption Eq.~(\ref{eq:stationarity}) the ``deterministic
approximation'' as it neglects stochastic fluctuations in the total population
density.

With Eq.~(\ref{eq:stationarity}), all parameters in the Fokker-Planck equation,
Eq.~(\ref{eq:FPE}), of the ancestral distribution $G(x,t|\xi,\tau)$ are
time-independent and its analysis considerably simplifies: If a unique stationary
solution $\gst(\xi)\equiv\lim_{t\to\infty}G(\xi,\tau|x,t)$ exists, it can be written
explicitely in terms of the stationary density profile $\cst(\xi)$,
\begin{equation}
  \label{eq:fixprob}
  \gst(\xi)\propto \cst^2(\xi)\exp\left( v \xi/D  \right) 
\end{equation}
where a pre-factor is required to satisfy the normalization condition of $\gst$,
$\int \gst(\xi,\tau|x,t) \,d\xi=1$.\footnote{In fact, strict stationarity,
  Eq.~(\ref{eq:stationarity}), is not necessary for Eq.~(\ref{eq:fixprob}) to hold,
  rather the drift coefficient in the Fokker--Planck equation
  (\ref{eq:comoving-Rdiff}) has to be time-independent.  This condition is satisfied
  whenever the density profile is separable, i.e., $c(x,t)=g(t)h(x)$, for two
  functions $g(t)$ and $h(x)$. } 

As shown below, the analytical expression Eq.~(\ref{eq:fixprob}) describes at least
qualitatively the bell-like shapes found for the ancestral distribution function
$\avg{P_i}$ in our stochastic simulations. The exponential factor biases the fixation
probability~\footnote{The actual fixation probability~\cite{RefWorks:33} $u(\xi)$ of
  a mutation occurring in a \emph{single} individual at $\xi$ is obtained from $\gst$
  after dividing by the population density, $u(\xi)=\gst(\xi)/\cst(\xi)$. This
  quantity measures the probability of ultimate evolutionary success of a neutral
  genetic marker in a single individual at location $\xi$ in the wave front.} towards
the tip of the wave ($\xi>0$) and competes with the pre-factor controlled by the
decaying density of individuals in the tip of the wave.

It is noteworthy that Eq.~(\ref{eq:fixprob}) not only applies to range expansions,
but can be evaluated for any deterministic population dynamics, such as deterministic
models of evolution~\cite{tsimring96,Rouzine:W:C::100:p587-592:2003} (where however
rare events might be crucial as found in Ref.~\cite{munier06}) and to source-sink
populations~\cite{RefWorks:34,RefWorks:22}, a simple example of which is given in the
Appendix \ref{app:b}. If the spatial domain is unbounded, Eq.~(\ref{eq:fixprob}) yields finite
results as long as $\cst(\xi)$ decays faster than $\exp[-v \xi/(2D)]$ as
$\xi\to\infty$.  This condition formally distinguishes the two classes of waves
earlier denoted by {\pulled} and {\pushed}. Within the mean-field description of such
waves, the right hand site of Eq.~(\ref{eq:fixprob}) is normalizable only in the case
of {\pushed} waves~\cite{ebert00}. The density of {\pulled} waves, however, decays as
$\exp[-v \xi/(2D)]$ in the foot of the wave ($\xi\to\infty$) as follows from a
linearized mean field treatment~\cite{RefWorks:23}. The prime example of {\pulled}
waves, the mean-field Fisher wave, does not therefore allow for successful surfing,
$\gst\equiv 0$, as is explicitely shown in Appendix \ref{app:c}. This is in marked contrast to our
simulations of \emph{stochastic} Fisher waves (Fig.~\ref{fig:exple}a). There, we
found finite bell-like ancestral distributions up to deme sizes on the order of
$10^5$.  This striking discrepancy indicates that the classical Fisher equation is a
poor approximation for the case of finite deme sizes (even if they are large).  An
improved deterministic equation with a modified reaction term has been
proposed~\cite{brunet97}, which is able to reproduce the leading reduction in wave
velocity due to the discreteness.  A remarkable property of Eq.~(\ref{eq:fixprob}) is
that it should be valid, \emph{irrespective} of the actual form of the reaction term,
if the demography is deterministic. By comparing Eq.~(\ref{eq:fixprob}) to
simulations, it is possible to test the deterministic character of a population wave,
i.e., whether or not a deterministic reaction-diffusion description might be
appropriate.

\begin{figure}
    \includegraphics{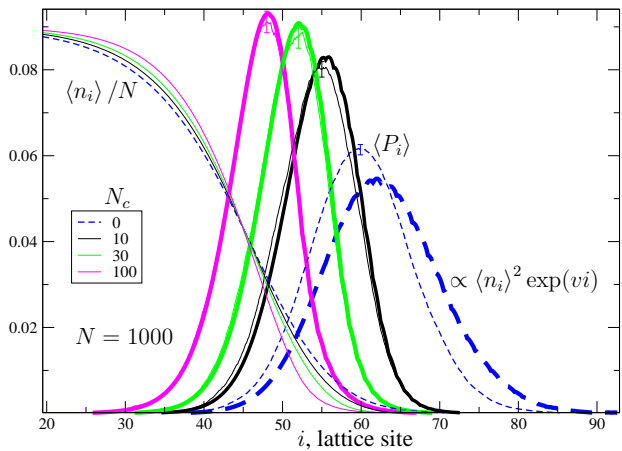}
    \caption{ The data from Fig.~\ref{fig:exple}b (thin lines) superimposed with the prediction
      Eq.~(\ref{eq:fixprob}) (thick lines) for the ancestral distribution function
      $\avg{P_i}$ based on the measured wave profile $\avg{n_i}$ and velocity $v$.
      As explained in the text, the apparent systematic deviations in the {\pulled}
      case ($N_c=0$) are caused by fluctuations in the tip of the wave.}
  \label{fig:theory}
\end{figure}

\begin{figure*}
  \includegraphics{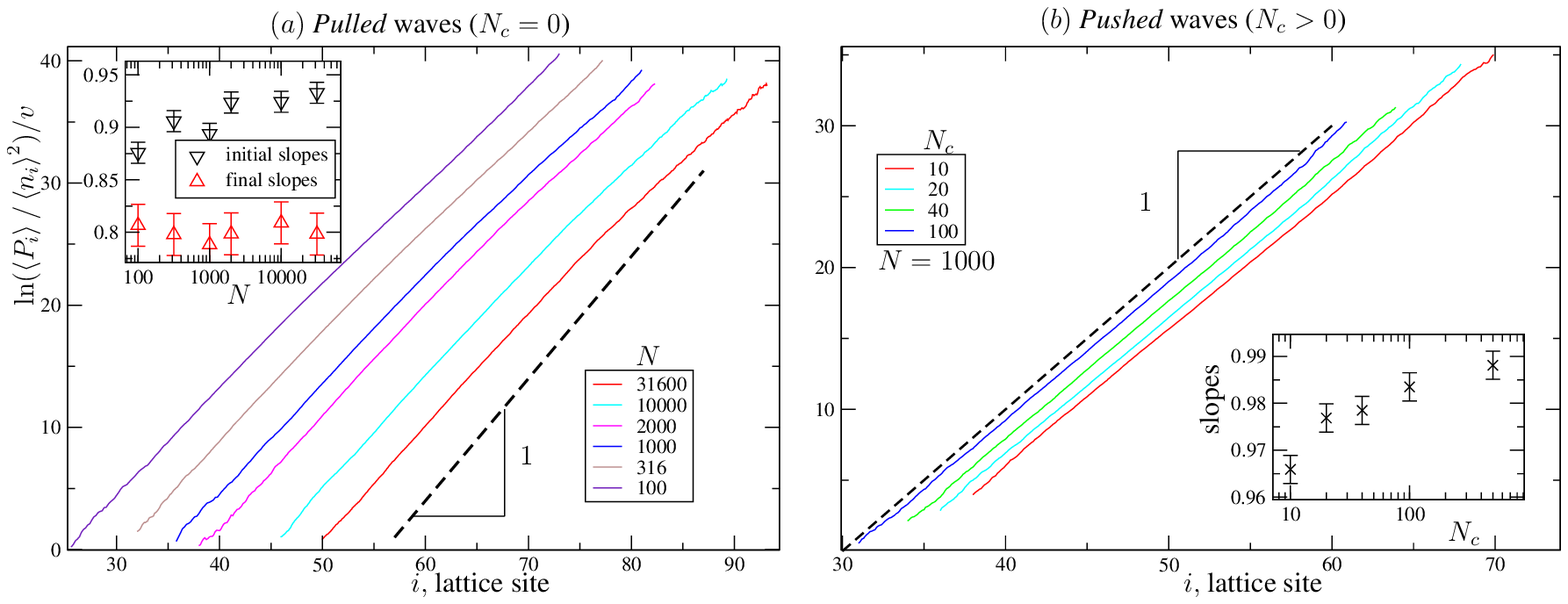}
  \caption{The quantity $\ln(\avg{\gex_i}/\avg{\cex_i}^2)/v$ as a function of lattice
    site $i$ in the co-moving frame, which should have slope $1$ according to the
    deterministic approximation, Eq.~(\ref{eq:fixprob}). (a) Results for {\pulled}
    waves ($N_c=0$) with various deme sizes $N$; the values for the ``initial slope''
    and ``final slope'' have been obtained by fitting a straight line to the lower
    and upper half of each shown curve, respectively.  (b) Results for various
    {\pushed} waves ($N_c>0$) with fixed deme size, $N=1000$. (The domain of each
    curve is restricted to a region, in which the bell-like distributions
    $\avg{\gex_i}$ has enough support to sample a sufficient amount of data points.
    This region roughly covers $98\%$ of all successful surfing events.) }
  \label{fig:log-lin}
\end{figure*}

For our simulations, such a test is given in Fig.~\ref{fig:theory}, where we
superimpose measured ancestral distribution functions $\avg{P_i}$ with those
predicted by Eq.~(\ref{eq:fixprob}) based on the measured wave velocity $v$ and the
occupation numbers $\avg{n_i}$ (the discrete analog of the population density
$\cst(\xi)$).  It is seen that systematic deviations occur in the {\pulled} case
($N_c=0$), where the predicted distribution seems to be somewhat displaced towards
the tip of the wave.  The agreement of theory and simulation is much better for
$N_c=10$ and further improves when $N_c$ is increased.  Altogether, our deterministic
approximation Eq.~(\ref{eq:fixprob}) seems to apply best to {\pushed} waves with a
strong Allee effect ($N_c\gg1$), whereas significant deviations to
Eq.~(\ref{eq:fixprob}) occur for {\pulled} waves. An alternative test of theory and
simulation, presented in Fig.~\ref{fig:log-lin}, supports this conclusion and furthermore
shows that the deterministic approximation applied to {\pulled} waves improves slowly
with increasing deme size.  For reasonable system sizes, however, fluctuation effects
in {\pulled} waves are non-negligible~\cite{van03}.

\subsection{Decrease of genetic diversity} To measure how fast genetic diversity
decreases at the wave front due to gene surfing, we also studied in our simulations
$\hex(t)$, defined in Eq.~(\ref{eq:Phi}) as the probability that two randomly sampled
individuals carry different labels at a time $t$ after a labeling event. To what
extent are the decrease of $\hex(t)$ and the shape of $\gex(\xi)$ related? As before,
a retrospective view on the problem simplifies the theoretical analysis.  Imagine
following the lineages of two randomly sampled individuals backward in time.  They
will drift and diffuse separately for a certain time $t_c$ until they \emph{coalesce}
in the most recent common ancestor.  If the last labeling event occurred at an earlier
time $t>t_c$ (reversed time-direction) then both individuals must carry identical
labels.  If, on the other hand, $t<t_c$ then these individuals will have different
labels unless their different ancestors happen to be in the same deme at the labeling
time.  Up to a small error of the order of the inverse size of the simulation box, we
may thus identify the probability $\hex(t)$ of two individuals carrying different
labels with the probability that their coalescence time $t_c$ is larger than $t$.

In principle, the coalescence time distribution of two lineages can be explored by
studying the simultaneous backward--diffusion process of two lineages conditional on
having not coalesced before~\cite{Wilkins:W::161:p873-888:2002}. This process is
described by a generalization of Eq.~(\ref{eq:comoving-Rdiff}) augmented by a
well-known sink term~\cite{Wilkins:W::161:p873-888:2002} that accounts for the
probability of coalescence when lineages meet.

Here, we describe a (more tractable) approximation that estimates the behavior of
$\hex(t)$ from the distribution $\gex(\xi)$, analyzed in the previous section.  It is
based on the assumption that the coalescence rate of two lineages is so small that
each lineage has enough time to equilibrate its spatial distribution before
coalescence occurs.  Under this quasi-static approximation, the behavior of $\hex(t)$
is described by~\cite{nagylaki80}
\begin{equation}
  \label{eq:ctime}
  \partial_\tau \hex\approx-2 \hex \int \frac{\gex^2(\xi,\tau)}{c(\xi,\tau)} \,d\xi\;,
\end{equation}
when time is measured in units of generation times. The justification of
Eq.~(\ref{eq:ctime}) is as follows: The coalescence rate, $-\partial_\tau \hex$, at
time $\tau$ in the past is given by the probability that the two lineages have not
coalesced earlier, $\hex(\tau)$, times the rate at which two separate lineages
coalesce at time $\tau$.  The latter is locally proportional to the product of the
probabilities that the two lineages meet at the same place, $\propto P^2(\xi,\tau)$,
and that they meet in the same individual, $\propto c^{-1}$, given they are at the
same place. Less obvious, unfortunately, is the numerical pre-factor ``$2$'' on the
right-hand side, which is specific to the employed breeding scheme (Moran
model\cite{RefWorks:33}
).

Eq.~(\ref{eq:ctime}) yields the correct coalescence time distribution in the
so-called \emph{strong migration limit}~\cite{nagylaki80,NOTOHARA::31:p115-122:1993}
of large population densities, $c\to \infty$, while the diffusivity $D$ and the
spatial extension of the habitat are held fixed. In our case, it serves as a simple
approximation that tends to overestimate coalescence rates, because it neglects
spatial anti-correlations between non-coalescing lineages: Lineages that have avoided
coalescence will usually be found further apart than described by the product of
(one-point) distribution functions in Eq.~(\ref{eq:ctime}). Thus, their rate of
coalescence will, typically, be smaller than in Eq.~(\ref{eq:ctime}).

Equation~(\ref{eq:ctime}) predicts exponential decay, $\hex(t)\sim \exp[-2
(\tau-t)/\nef]$, in the deterministic approximation, Eq.~(\ref{eq:stationarity}),
with a rate depending on a constant $\nef$ given
by~\cite{nagylaki80}
\begin{equation}
  \label{eq:effective-popsize}
  \nef^{-1}= \int \frac{\gst^2(\xi)}{\cst({\xi})}\,d{\xi}\;.
\end{equation}
In fact, a generalization of this argument to the coalescence process of a sample of
$n$ lineages shows that the \emph{standard
  coalescent}~\cite{KINGMAN::13:p235-248:1982} is obtained in the strong-migration
limit with the parameter $\nef$ interpreted as the \emph{effective population
  size}
~\cite{nagylaki80}.  In other words, the coalescence
process in the strong-migration limit is identical, in every respect, to the
coalescence of a well-mixed population of fixed size $N_e$.

The strong-migration approximation may be tested by comparing the effective
population sizes measured in our simulations with the ones predicted by
Eq.~(\ref{eq:effective-popsize}) based on the measured ancestral distribution
$\avg{\gex_i}$ and the number density profile $\avg{\cex_i}$. These inferred values
are plotted in Fig.~\ref{fig:log-log-effpopsizes} as (red) triangles. For both
{\pushed} and {\pulled} waves, the agreement between inferred and measured effective
population sizes becomes excellent for large deme sizes, $N>100$. For lower values of
$N$ the strong migration assumption overestimates genetic drift, presumably due to
the neglect of correlations as mentioned above.

\section{Discussion}
\label{sec:discussion}

We have studied the impact of a range expansion on the genetic diversity of a
population by means of simulations and analytical techniques. The one dimensional
case treated in this article applies to populations following a (possibly curved)
line, like a  migration route, coast line, river or railway track. We have further
simplified our analysis by neglecting habitat boundaries, which is appropriate for
describing the colonization period, as long as the wave front is sufficiently far
away from the boundaries.

Our findings suggest the following general scenario. Suppose, an initially well-mixed
population increases its range from a smaller to a larger habitat, and that mutations
may be neglected on the time scale of the range expansion. Our simulations show that
the heterozygosity at the moving colonization front decays, due to genetic drift,
exponentially in time with a rate $2/N_e$, depending on an {\it effective} population
size $N_e$.  Upon combining this rate with the velocity $v$ (per generation) of the
colonization front, we obtain a length $\lambda=v N_e/2$ that characterizes the
pattern of genetic diversity generated by the colonization process: As the wave front
moves along it leaves behind saturated demes with heterozygosities given by the value
of the front heterozygosity as the wave front passes through.  The transient
colonization process therefore engraves a spatially decreasing profile of
heterozygosity into the newly founded habitat.  This profile decays exponentially in
space on the characteristic length $\lambda$ and serves as an initial condition for
the succeeding period of demographic equilibration, which may be described by
traditional models of population genetics~\cite{RefWorks:33}.

Of critical importance for the interpretations of gene frequency clines in natural
populations is the question as to which parameters chiefly control $\lambda$. Our
computer simulations have revealed, that the effective population size, and thus
$\lambda$, only grows \emph{sub-linearly} with increasing deme size, in contrast to
the naive expectation that the effective population size should roughly be given by
the characteristic number of individuals contained in the wave front (deme size times
the width of the wave).  On the other hand, we found that the population
``bottleneck'' at the wave front was significantly widened when we implemented an
Allee effect~\cite{Allee31} into our model, by which growth rates for small
population densities are decreased. As a consequence, the region of major growth
shifted away from the frontier into the bulk of the wave and, thus, the effective
size of the gene pool for the further colonization was increased.

Finally, we have developed a theoretical framework to study the backward-dynamics of
neutral genetic markers for a given non-equilibrium population dynamics, which is
summarized by the Fokker-Planck equation (\ref{eq:FPE}) and its generalization in
Appendix \ref{app:a}. In stationary populations, it leads to a simple expression,
Eq.~(\ref{eq:fixprob}), for the long-time probability distribution of the common
ancestors, which can be used, in the strong-migration limit to determine the
effective population size, via Eq.~(\ref{eq:effective-popsize}). Comparison with our
stochastic simulations reveals that the simple deterministic results are good
approximations to stochastic simulations in the case of {\pushed} waves (strong Allee
effect), but that significant deviations occur for {\pulled} waves due to the
fluctuations in the frontier of the population wave. For sufficiently large deme
sizes, the effective population size could, in both cases, be inferred with
remarkable accuracy from Eq.~(\ref{eq:effective-popsize}).

\begin{acknowledgments}
  This research was supported by the German Research Foundation through grant no.~Ha
  5163/1 (OH). It is a pleasure to acknowledge conversations with Michael Brenner,
  Michael Desai and John Wakeley.
\end{acknowledgments}

\appendix

\section{Basic formalism}
\label{app:a}
In this appendix, we describe the derivation of the Fokker--Planck equation, Eq.~(\ref{eq:FPE}),
and its generalization to heterogeneous migration and higher dimensions.

Central to our analysis is the assumption that the total population consists of
statistically identical entities, such that two different individuals at a given time
and location behave in the same way. In particular, we assume that migration as well
as reproduction of an individual are \emph{exchangeable}~\cite{RefWorks:35} random
processes, i.e., independent of any label that might be assigned to it.

Let us indeed imagine a subpopulation labeled by a neutral marker, and ask: What is
the dynamics of the labeled individuals for a \emph{given} dynamics of the total
population, as described by Eq.~(\ref{eq:comoving-Rdiff})? It is clear that the density $c^\star( x ,t)$ of
labels obeys a reaction-diffusion equation with coefficients $D^\star$, $v^\star$ and
$K^\star$ that are closely related to the one for total population by means of the
exchangeability assumption.  Firstly, labeled and unlabeled individuals should
migrate statistically in the same way, which is measured by diffusion and drift
coefficients, i.e., we have $D^\star=D$ and $v^\star=v$.  Secondly, the labeled
subpopulation must carry the number fluctuations of the total population, encoded in
$K$, in proportion to its reduced size: $K^\star=K c^\star/c$.  However, these
statements are true only \emph{on average}: The discreteness of the particle numbers
lead to fluctuations, for instance, in the quantity $K^\star-(c^\star/c)K$. To
illustrate this point, imagine that at a given time, the term $K$ dictates that an
individual dies in some small spatio-temporal region, then this individual has to be
sampled from the labeled subpopulation with probability $c^\star/c$. The fluctuations
of $c^\star$ due to this sampling procedure represents random genetic drift and must
have zero mean according to the exchangeability assumption.  Similar fluctuations
affect the migration currents of the labeled subpopulation.  Thus, only upon
averaging over this source of stochasticity, we may formulate a reaction-diffusion
equation of the form
\begin{equation}
  \label{eq:eom-star-simple}
  \partial_t \overline{c}^\star=D\partial_x^2\overline{c}^\star+ v\partial_x
  \overline{c}^\star +K  \, \frac{\overline{c}^\star}{c}
\end{equation}
for the average density $\overline{c}^\star( x,t)$ of labeled individuals. In
Eq.~(\ref{eq:eom-star-simple}), only the effect of the genetic drift of labeled
individuals within the total population has been averaged out, the number
fluctuations affecting the total population density are retained through the
fluctuating reaction term $K$. In other words, Eq.~(\ref{eq:eom-star-simple})
describes the behavior of labeled individuals, if we average over many realizations
conditional on a given fixed evolution of the total population, described by
Eq.~(\ref{eq:comoving-Rdiff}).  Note that, this averaging ``works'' because
Eq.~(\ref{eq:eom-star-simple}) is linear in $c^\star$, such that a noise term with
zero mean added to the right--hand--side to account for the genetic drift can be
averaged out without generating higher moments.

Next, we use \[ K(x,t)=\partial_t c(x,t)-D\partial_x^2 c(x,t)-v\partial_x c(x,t)\]
from Eq.~(\ref{eq:comoving-Rdiff}) to substitute the reaction term $K$ in
Eq.~(\ref{eq:eom-star-simple}). After rewriting the average density
$\overline{c}^\star(x,t)\equiv\overline{p}(x,t) c( x,t)$ of labeled individuals in
terms of their average frequency (or ratio) $\overline{p}$, we obtain
\begin{equation}
  \label{eq:eom-p-simple}
  \partial_t \overline{p}( x,t)=D\partial_x^2\overline{p}
  +\left\{v+2 D\partial_x \ln[c(x,\tau)] \right\}\partial_x \overline{p} \;.
\end{equation}
Notice that, $\overline{p}=$const.~is a (steady state) solution of
Eq.~(\ref{eq:eom-p-simple}). Relations formally equivalent to
Eq.~(\ref{eq:eom-p-simple}) have been formulated in Refs.~\cite{nagylaki75,
  RefWorks:36} in different contexts, in which, unlike the present case, random
genetic drift could not be averaged out due to non-linearities, but had to be
disregarded, instead. As a transport equation for \emph{deterministic} gene
frequencies, an equation similar to Eq.~(\ref{eq:eom-p-simple}) was also recently
obtained in Ref.~\cite{VladCR04}.

For a given realization of the time-evolution of the total population, the quantity
$\overline{p}(x,t)$ can be interpreted as the probability that an individual sampled
from $( x,t)$ is labeled, i.e., that it has descended from the initially labeled
population. If the above tracer experiment for a \emph{given} dynamics of $c( x,t)$
is repeated multiple times, $\overline{p}( x,t)$ represents the histogram of the
number of times a individual sampled from $( x,t)$ is labeled.

By choosing proper initial conditions, this allows us to study ``where individuals
come from'': Suppose that, at time $\tau$, the labeled population contains all
individuals within a small interval around the position $\xi$. The solution of
Eq.~(\ref{eq:eom-p-simple}) for later times will then tell us the probability that a
individual at $( x,t)$ has descended from an ancestor sampled from that narrow region
around $\xi$ at time $\tau$.

Hence, the probability density $G( \xi,\tau|x,t)$, introduced above Eq.~(\ref{eq:FPE}), that an
individual at $(x,t)$ has an ancestor who lived at $(\xi,\tau)$, is just the solution
of Eq.~(\ref{eq:eom-p-simple}) for the initial condition
\[ G(\xi,0|x,0)=\delta( x- \xi) \;,\] where $\delta( x)$ is the Dirac delta function.
It is straightforward~\cite{vankampen01} to show that this Green's function of
Eq.~(\ref{eq:eom-p-simple}) also obeys the Fokker--Planck equation (\ref{eq:FPE}), in which
time is measured in the backward direction; Eq.~(\ref{eq:eom-p-simple}) is usually
called the (Kolmogorov--) backward equation~\cite{vankampen01} associated with the
Fokker-Planck equation (\ref{eq:FPE}).

The content of the Fokker-Planck equation (\ref{eq:FPE}) may be further illustrated by a
physical analog.  The dynamics of a single ancestral line backward in time
conditional on a particular demographic history, as described by Eq.~(\ref{eq:comoving-Rdiff}), is
equivalent to the Brownian motion of a particle (with $k_B T=1$) in a potential
$U(\xi,\tau)$, whose negative gradient is given by the drift coefficient $v+2
D\partial_\xi \ln(c)$ in Eq.~(\ref{eq:FPE}).  The form of this time-dependent potential,
\begin{equation}
  \label{eq:potential}
  U(\xi,\tau)=-v\xi-2D\ln[c(\xi,\tau)] \;,
\end{equation}
suggests an interpretation in terms of a fluctuating free energy in which the first
and second term are the energetic and entropic contribution, respectively. If the
spatial domain is unbounded, the long-time distribution function of the fluctuating
particle will be non-trivial only if the potential Eq.~(\ref{eq:potential}) has the
form of a well, which is able to ``trap'' the fluctuating particle on long times.
Otherwise, for instance if the potential is half-open as in the case of a Fisher
wave, the ancestral probability distribution will decay to zero on long times.

So far, our analysis was restricted to a one-dimensional reaction diffusion model, in
which drift and diffusion are space-independent.  The generalization of Eq.~(\ref{eq:comoving-Rdiff}) to
heterogeneous migration and higher dimensions reads
\begin{equation}
  \label{eq:eom}
  \partial_t c=\frac{\partial}{\partial x_i} \left(D_{ij}\frac{\partial c}{\partial x_j} \right)+
  \frac{\partial}{\partial x_i}  \left(v_i c\right)+K \;.
\end{equation}
The conservative dynamics is now described by a matrix $D_{ij}(\vec x,t)$ of
diffusivities and a velocity vector $v_i(\vec x,t)$, which may both depend on space
and time. Again, the reaction term $K(\vec x,t)$ accounts for deterministic and
stochastic fluctuations in the number of individuals due to birth and death
processes.

Repeating the above arguments for the dynamics of neutral labels under the more
general population dynamics Eq.~(\ref{eq:eom}) yields a multi-dimensional
Fokker--Planck equation for the probability density $G(\vec \xi,\tau|\vec x,t)$ that
an individual at $(\vec x,t)$ has descended from an ancestor who lived at $(\vec
\xi,\tau)$, which is given by
\begin{eqnarray}
  \label{eq:FPE-app}
  \partial_\tau G(\vec \xi,\tau|\vec x,t)&=&-\frac{\partial}{\partial \xi_i}
  J_i(\vec \xi,\tau|\vec x,t) \\
  J_i(\vec \xi,\tau|\vec x,t)&\equiv &-\frac{\partial}{\partial \xi_j} (D_{ij}
  G)+\nonumber\\
  & &\left[v_j-\frac{\partial D_{ij}}{\partial \xi_j}+\frac{2}{c} \frac{\partial(D_{ij}c)}{\partial \xi_j} \right] G \nonumber  \;.
\end{eqnarray}
Here, summation over identical indices is implied and time again increases in the
backward direction.  The natural requirement that there is no probability flux $\vec
J$ out of the region $S$ of non--vanishing population density leads to reflecting
boundary condition, $J=0$, on the boundary $\partial S$ of $S$.  Note that our
stochastic description of the backward dynamics of a single lineage,
Eq.~(\ref{eq:FPE-app}), is fully determined by the demographic history $c(x,t)$.  A
knowledge of the actual form of the reaction term $K(x,t)$ in the reaction-diffusion
equation (\ref{eq:eom}) is not necessary.

\section{Source-sink populations}
\label{app:b}
For purely conservative populations~\cite{nagylaki00} of neutral individuals, subject
only to diffusion and drift, it is well-known that the fixation probability is the
same for all individuals, $ u=$const.$=1/N$, and that the effective population size
equals the total population, $N_e=N$. However, when reaction terms are important,
individuals become privileged or handicapped depending on where they linger. As a
telling example, let us consider the case of an ``oasis''~\cite{RefWorks:22} (or
source~\cite{Gaggiotti::50:p178-208:1996}) with a carrying capacity $c_o$ in
equilibrium with a ``desert'' (a sink) of smaller carrying capacity $c_d$.
Sufficiently far away~\footnote{The transition zone from oasis to desert has a
  characteristic width $\sqrt{D/s}$ that depends on a diffusion constant and the
  effective growth rate $s$.}  from the contact zone, the population densities will
be saturated at their respective carrying capacities. According to Eq.~(\ref{eq:fixprob}), the
ancestral distribution function of a stationary non-moving population will be locally
proportional to the square of the density, $\gst(\xi)\propto \cst^2(\xi)$. As noted
in footnote 5 of the main text, the fixation probability of a mutation occurring in a
single individual is given by the ratio of ancestral distribution function and
population density, $u(\xi)\propto \gst(\xi)/\cst(\xi)$.  Thus, the probability that
a neutral mutation fixes will be larger if it arises deep in the oasis, $u_o$, than
if it arises in the desert, $u_d$. The ratio of both fixation probabilities is given
by the ratio of the respective densities,
\begin{equation}
  \label{eq:ratio-of-fprobs}
  \frac{ u_o}{ u_d}= \frac{c_o}{c_d} \;.
\end{equation}
Apart from its simplicity, the relation Eq.~(\ref{eq:ratio-of-fprobs}) is remarkable
because it is independent of the diffusion constants and the details of the
particular logistic interaction between individuals.  

If the combined system of oasis and desert is closed, the effective population size,
Eq.~(\ref{eq:effective-popsize}), evaluates to
\begin{equation}
  \label{eq:ne-oasis-desert}
  \nef=\frac{(c_d^2 L_d+c_o^2 L_o)^2}{c_d^3 L_d+c_o^3 L_o} < c_d
  L_d+c_o L_o
\end{equation}
to leading order in the linear sizes $L_o$ and $L_d$ of oasis and desert. As
mentioned earlier, our reasoning regarding coalescence times only applies to the
strong--migration limit, in which the fixation time $\sim \nef T$, where $T$ is the
generation time, is much smaller than the longest relaxation time of the
Fokker--Planck equation. For the present case, the latter may be estimated by the
time needed for lineage to cross the habitat, $\sim L^2/D$.

\section{Surfing on a Fisher wave}
\label{app:c}
Here, we apply our theory of gene surfing to the 
Fisher equation~\cite{fisher37}, 
\begin{equation}
  \label{eq:fk-wave-eq}
  \partial_t c(x,t)=D\partial_x^2 c(x,t)+s\left[ 1-\frac{c(x,t)}{c_\infty}  \right]c(x,t) \;,
\end{equation}
which is the prime example of a {\pulled} front.  This equation was originally
proposed as a mean--field model for the spread of a dominant gene with selective
fitness advantage $s$ through a population with constant density $c_\infty$ (carrying
capacity).  Equation (\ref{eq:fk-wave-eq}) has also been useful as a description of
an expanding population, for which $s$ is the difference between the linear birth and
death rates, and the term $-s c^2/c_\infty$ represents some self-limiting process,
roughly proportional to the number of pairs of individuals at position $x$.

There are two spatially homogeneous fixed points: an unstable fixed point at
$c(x)=0$, in which there is no population at all, and a stable fixed point at
$c(x)=c_\infty$, where the population saturates to the carrying capacity $c_\infty$
of the environment.

Non-negative initial configurations evolve smoothly toward the stable fixed point;
analysis of the time development of spatial fluctuations in this model reveals that
equilibrium can be reached via traveling soliton-like solutions $\cst(x-v t)$,
referred to as Fisher waves~\cite{fisher37}, which represent steady state solutions
of Eq.~(\ref{eq:fk-wave-eq}). In the wave front, where the population density is much
smaller than the carrying capacity, the non-linear logistic term $\propto c^2$ in
Eq.~(\ref{eq:fk-wave-eq}) may be neglected. The steady state solution of the
remaining linear equation is exponentially decaying,
\begin{equation}
  \label{eq:wprofile-tip}
  \cst\sim \exp(-x/\lambda)\;,
\end{equation}
for $x\to\infty$, where decay length $\lambda$ and velocity $v$ are related by
\begin{equation}
  \label{eq:velocity-dispersion}
  0=\frac{D}{\lambda^2}-\frac{v}{\lambda}+s \;.
\end{equation}
The population density to be nonnegative requires real solutions $\lambda>0$ of
Eq.~(\ref{eq:velocity-dispersion}), which do not exist unless
\begin{equation}
  \label{eq:velocity-disp2}
  v\geq 2\sqrt{D s}\;.
\end{equation} 
It can be shown that the solution corresponding to the lowest velocity $v=2\sqrt{D
  s}$ and decay length $\lambda= \sqrt{D/s}$ is approached for any initial conditions
with compact support, and, thus, is the solution most relevant for biological
applications~\cite{RefWorks:23}.

Now, when we evaluate the surfing probability, Eq.~(\ref{eq:fixprob}) for this standard model of a
spreading wave, we find zero - a somewhat surprising result in light of the finite
bell-like curves obtained from our simulations of stochastic Fisher waves (Fig.~2a).
The exponential decrease of the population density at the wave front,
Eqs.~(\ref{eq:wprofile-tip}, \ref{eq:velocity-disp2}), is simply not fast enough to
render the function $\cst^2(\xi)\exp(v \xi/D)$ normalizable - not even for the lowest
velocity for which it approaches a positive constant as $\xi\to\infty$.  Hence, a
non--zero stationary distribution function of the common ancestor does not exist,
even though the total population is in a steady state.

A closer look to the \emph{dynamics} of a lineage, as described by the Fokker--Planck
equation (\ref{eq:FPE}), reveals how the distribution of the common ancestor decays to zero
with time. As we evolve the probability distribution $G(\xi,\tau|x,t=0)$ that a
lineage diffuses from a location $x$ to $\xi$ back through time, it spreads out, due
to diffusion, and is subject to a drift of strength $v+2D\partial_\xi \log \cst$. If
a lineage starts out deep in the wake of the wave, $x\ll-1$, where $\partial_\xi \log
\cst\to 0$, it experiences a drift pushing it towards the wave front. After a time of
the order of $\abs{x}/v$, the probability cloud of the single lineage reaches the
front and, when the inflection point is passed, the drift has decreased appreciable
because $-\partial_\xi \log \cst=0(1)$.  In the tip of the wave, the density profile
is exponential, Eq.~(\ref{eq:wprofile-tip}), such that the drift,
$v+2D\partial_\xi\log(\cst)$, saturates at a \emph{non-negative} value $w\equiv\left(
  v-2D/\lambda \right)\geq 0$, which tends to move the lineage even further away from
the bulk into the tip of the wave.  For large times, the distribution $G$ assumes the
form of a bell--like curve of width $\sim \sqrt{D \tau}$ moving with a velocity $w$.
For any fixed $\xi$ in the foot of the wave and $w>0$, the distribution function $G$
decays exponentially to zero for times, when the probability ``cloud'' has passed
$\xi$ .  Only for $w=0$, which corresponds to the lowest allowed velocity $v=2\sqrt{s
  D}$, drift is absent in the tip of the wave, such that the decay is much slower,
$G\to \tau^{-1/2}$. In any case, however, $G$ decays to zero for any location $\xi$,
which means that, no matter which individual we choose, the fixation probability is
zero.  Thus, ``successful surfing'' is not possible in the case of the deterministic
Fisher wave, as was mentioned already in the main text. We expect that, in a {\it
  stochastic} simulation of a finite number of particles, the time--dependent
features discussed above for the mean-field Fisher equation are merely {\it
  transient} and only visible for times smaller than the longest relaxation time of
the Fokker--Planck equation, Eq.~(\ref{eq:FPE}). This conjecture can be supported by employing
an approximation scheme due to Brunet-Derrida~\cite{brunet97} to take into account
leading order effects of discreteness.


\begin{thebibliography}{10}








\bibitem{templeton2002oaa} Templeton, A. (2002) {\it Nature} {\bf 416}, 45-51.

\bibitem{RosenbergPWCKZF03} Rosenberg, N.~A., Pritchard, J.~K., Weber, J.~L., Cann,
  H.~M., Kidd, K.~K., Zhivotovsky, L.~A., and Feldman, M.~W. (2002) {\it Science}
  {\bf 298} (5602), 2381 - 2385.

\bibitem{RamachandranDRRFC05}
Ramachandran, S., Deshpande, O., Roseman, C.~C., Rosenberg,  N.~A., Feldman,  M.~W.,
  and Cavalli-Sforza,  L.~L. (2005) {\it Proc. Natl. Acad. Sci. USA}, 15942-15947.

\bibitem{cavallisforza93} Cavalli-Sforza, L.~L., Menozzi, P., and Piazza, A. (1993)
  {\it Science} {\bf 259}, 639--646.



\bibitem{phillips2006iae}
Phillips, B.~L., Brown, G.~B., Webb, J.~K., and Shine, R. (2006)
{\it Nature} {\bf 439} (7078),803.


\bibitem{Hewitt00}
Hewitt, G.~M., (2000) {\it Nature} {\bf 405}, 907-913.

\bibitem{currat06}
Currat, M., Excoffier, L., Maddison, W., Otto, S., Ray, N., Whitlock, M.~.C., and
Yeaman, S. (2006) {\it Science} {\bf 313}, 172a.



\bibitem{RefWorks:32}
Hewitt, G.~M., (1996) {\it Biol. J.  Linn. Soc.} {\bf 58}(3),247-276.


\bibitem{RefWorks:19} Rousset, F., (2001) {\it Handbook of Statistical Genetics}
  (John Wiley \& Sons, London), 239-269.

\bibitem{RefWorks:33}
Hartl, D.~L.  and Clark, A.~G.  (1997) {\it Principles of population genetics}
Sinauer Associates Sunderland, Mass..

\bibitem{nichols94}
Nichols, R.~A.  and Hewitt,  G.~M. 
(1994) {\it Heredity} {\bf 72},312-317.

\bibitem{AusterlitzG03}
Austerlitz, F., and Garnier-Gere,  P.~H. (2003) {\em Heredity} {\bf 90}, 282-290.

\bibitem{EdmondsLC04} Edmonds, C.~A., Lillie, A.~S., and Cavalli-Sforza, L.~L. (2004)
  {\it Proc. Natl. Acad. Sci. USA} {\bf 101},975-979.

\bibitem{KlopfsteinCE06} Klopfstein, S., Currat, M. and Excoffier, L. (2006) {\it
    Mol. Biol.  Evol.} {\bf 23},482-490.

\bibitem{LiuPMB06}
Liu, H., Prugnolle,  F., Manica, A., and Balloux, F. (2006) {\it Am. J. Hum. Genet.}
{\bf 79},230-237.

\bibitem{AusterlitzJGG97}
Austerlitz, F.~, Jung-Muller, B., Godelle,  B., and Gouyon, P.~H.
(1997) {\it Theor. Pop. Biol.} {\bf 51}, 148-164.

\bibitem{LeCorreK98} Le~Corre, V.~ and Kremer, A. (1998) {\em J. Evol. Biol.} {\bf
    11}, 495-512.

\bibitem{Allee31} Allee, W.~C. (1931) {\it Am. J. Soc.} {\bf 37} (3), 386-398.

\bibitem{van03} van Saarloos, W. (2003) {\it Phys. Rep.}  {\bf 386}, 29-222.

\bibitem{panja04} Panja, D. (2004) {\it Phys. Rep.} {\bf 393}, 87-174.


\bibitem{courchamp99} Courchamp, F., Clutton-Brock, T., and Grenfell, B. (1999) {\em
    Trends Ecol. Evol.} {\bf 14}, 405-410.

\bibitem{RefWorks:30} Doering, C.~R., Mueller, C., and Smereka, P. (2003) {\it
    Physica A}, {\bf 325}(1-2), 243-259.

\bibitem{fisher37}
Fisher, R.~A. (1937) {\it Ann. Eugenics} {\bf 7}, 355-369.



\bibitem{vankampen01} van Kampen, N.~G. (2001) {\it Stochastic processes in physics
    and chemistry}.  (Elsevier, Amsterdam).

\bibitem{VladCR04} Vlad, M.~O., Cavalli-Sforza, L.~L., and Ross, J. (2004) {\it Proc.
    Natl. Acad. Sci. USA} {\bf 101}, 10249-10253.


\bibitem{BARTON:W::349:p49-59:1995} Barton, N.~H.  and Wilson, I. (1995) {\it Philos.
    Trans. R. Soc. London, Ser. B} {\bf 349}(1327), 49-59.


\bibitem{tsimring96}
Tsimring, L.~S., Levine,  H., and Kessler, D.~A.
(1996) {\it Phys. Rev. Lett.} {\bf 76}, 4440-4443.

\bibitem{Rouzine:W:C::100:p587-592:2003} Rouzine, I.~M., Wakeley, J., and Coffin,
  J.~M. (2003) {\it Proc.  Natl. Acad. Sci. USA} {\bf 100} (2), 587-592.

\bibitem{munier06} Brunet, E., Derrida B., Mueller A.~H., and Munier, S. (2006) {\it
    Europhys. Lett.} {\bf 76}, 1-7.

\bibitem{RefWorks:34} Pulliam, H.~R. (1988) {\it Am. Nat.} {\bf 132} (5), 652-661.

\bibitem{RefWorks:22} Dahmen, K.~A., Nelson, D.~R. and Shnerb, N.~M. (2000) {\it J.
    Math. Biol.} {\bf 41} (1), 1-23.

\bibitem{ebert00}
Ebert, U. and van Saarloos, W. (2000) {\it Physica D} {\bf 146}, 1-99.

\bibitem{RefWorks:23}
Murray, J.~D. (2004) {\it Mathematical Biology} (Springer, New York).

\bibitem{brunet97}
Brunet, E. and Derrida, B. (1997) {\it Phys. Rev. E} {\bf 56}, 2597-2604.

\bibitem{Wilkins:W::161:p873-888:2002} Wilkins, J.~F. and Wakeley, J. (2002) {\it
    Genetics} {\bf 161} (2), 873-888.

\bibitem{nagylaki80}
Nagylaki, T. (1980) {\it J. Math. Biol.} {\bf 9}, 101-114.

\bibitem{NOTOHARA::31:p115-122:1993}
Notohara, M. (1993) {\it J. Math. Biol.} {\bf 31}(2), 115-122.

 

\bibitem{KINGMAN::13:p235-248:1982}
Kingman, J.~F.~C. (1982) {\it Stoch. Proc. Appl.} {\bf 13}, 235-248.

\bibitem{RefWorks:35} Aldous, D.~J. (1985) {\em Lecture notes in mathematics} {\bf 1117}, 1-198.

\bibitem{nagylaki75}
Nagylaki, T. (1975) {\it Genetics} {\bf 80}, 595-615.

\bibitem{RefWorks:36}
Pease, C.~M., Lande, R. and Bull, J.J. 
(1989) {\it Ecology} {\bf 70} (6), 1657-1664.


\bibitem{nagylaki00} Nagylaki, T.  (2000) {\it J. Math. Biol.} {\bf 41}, 123-142.


\bibitem{Gaggiotti::50:p178-208:1996}
Gaggiotti, O.~E. (1996) {\it Theor. Pop. Biol.} {\bf 50}(2), 178-208.

\end{thebibliography}
\end{document}